\documentstyle[aps,preprint,epsf,amssymb,graphicx]{revtex}
\tightenlines
\draft
\begin{document}
\title{The evolution of the universe 
from noncompact Kaluza-Klein theory}
\author{$^1$Jos\'e Edgar Madriz Aguilar\footnote{
E-mail address: edgar@itzel.ifm.umich.mx}
and $^{2}$Mauricio Bellini\footnote{
E-mail address: mbellini@mdp.edu.ar}}
\address{$^1$Instituto de F\'{\i}sica y Matem\'aticas,
AP: 2-82, (58040) Universidad Michoacana de San Nicol\'as de Hidalgo,
Morelia, Michoac\'an, M\'exico.\\
$^2$Departamento de F\'{\i}sica, Facultad de Ciencias Exactas y Naturales,
Universidad Nacional de Mar del Plata and
Consejo Nacional de Ciencia y Tecnolog\'{\i}a (CONICET),
Funes 3350, (7600) Mar del Plata, Argentina.}

\vskip .2cm
\maketitle
\begin{abstract}
We develope a 5D mechanism inspired in
the Campbell's theorem, to explain the (neutral scalar field governed)
evolution of the universe
from a initially inflationary expansion
that has a change of phase towards a decelerated
expansion and thereinafter evolves
towards the present day observed accelerated (quintessential)
expansion.
\end{abstract}
\vskip .2cm                             
\noindent
Pacs numbers: 04.20.Jb, 11.10.kk, 98.80.Cq \\
\vskip .1cm

\section{Introduction}

In a cosmological context,
the energy density of scalar fields has been recognized to contribute
to the expansion of the universe\cite{jd}, and has been proposed to
explain inflation\cite{Guth}, as well as the presently observed
accelerated expansion\cite{we}.
The observed isotropy and homogeneity of the universe do not allow for the
existence of long-range electric and magnetic fields, but neutral
scalar fields can have non-trivial dynamics in an expanding
Friedmann-Robertson-Walker (FRW) 
universe. An attempt to confront the data with the predictions
for a minimally coupled scalar field with an a priori unknown potential
was made recently\cite{sta}.

The idea that matter in 4D can be explained from
a 5D Ricci-flat ($R_{AB}=0$) Riemannian manifold is a consequence of the
Campbell's theorem. It says that any analytic $N$-dimensional Riemannian
manifold can be locally embedded in a $(N+1)$-dimensional Ricci-flat
manifold. This is of great importance for establishing the generality of the
proposal that 4D field equations with sources can be locally embedded
in 5D field equations without sources\cite{wesson}.
In other words, 4D matter can be induced by a 5D apparent vacuum.
Campbell's theorem is closely related to Wesson's interpretation of
5D vacuum Eintein gravity\cite{e1,e2,e3}. In view of this, it would
be of interest to consider the embedding of 4D cosmological solutions
in 5D Ricci-flat spaces.
In the Wesson's theory [called Space-Time-Matter (STM) theory],
the extra dimension is not assumed to be compactified, which is a major
departure from earlier multidimensional theories where the cylindricity
conditions was imposed. In this theory, the original motivation for assuming
the existence of a large extra dimension was to achieve the unification of
matter and geometry, i.e. to obtain the properties of matter as a consequence
of the extra dimensions.
For example, an attempt
to understand inflation
[which is governed by the neutral scalar (inflaton) field],
from a 5D flat Riemannian manifold
was made in\cite{NPB}.
During inflation, the scale factor of the universe accelerates and this
acceleration is driven by the potential energy associated with the
self-interactions of a scalar field. However, Campbell's theorem implies
that all inflationary solutions can be generated, at least in principle,
from 5D vacuum Einstein gravity\cite{e4}.
But, could be
possible to develope a formalism to describe all
the evolution of the universe?

The aim of this work
consists to develope a 5D mechanism inspired in
the Campbell's theorem, to explain the (neutral scalar field governed)
evolution of the universe
from a initially inflationary (superluminical) expansion
that has a change of phase towards a decelerated (radiation and later matter
dominated) expansion and thereinafter evolves
towards the present day observed accelerated expansion
(quintessence)\cite{ps}.
Although Campbell's theorem relates $N$-dimensional theories to
vacuum $(N+1)$-dimensional theories, it does not establish
a strict equivalence between them\cite{e31}. It is therefore important to
determine when such theories are equivalent. Clearly, this is
a more severe restriction than embedability. Two notions of
equivalence that could be considered are dynamical equivalence and
geodesic equivalence. Dynamical equivalence would imply that the dynamics
of vacuum $N$-dimensional theories is included in a vacuum $(N+1)$-dimensional
theories. Alternatively, one may consider geodesic equivalence, in the
sense of Mashhoon {\em et al.}\cite{e4}. In this case the
$(3+1)$ geodesic equation induces a $(2+1)$ geodesic equation plus
a force (per unity of mass) term $F^C$
\begin{displaymath}
\frac{dU^C}{dS} + \Gamma^C_{AB} U^A U^B = F^C.
\end{displaymath}
As was demonstrated in \cite{e4},
for canonical metrics the requirement $F^C=0$
holds. Hence, in such that metrics
it is not really an extra assumption that the
motion is geodesic.

In this work we shall use the geodesic equivalence approach.\\

\section{Formalism}

To make it, we consider the canonical 5D metric\cite{new,PLB}
\begin{equation}\label{6}
dS^2 = \epsilon\left(\psi^2 dN^2 - \psi^2 e^{2N} dr^2 - d\psi^2\right),
\end{equation}
where $dr^2=dx^2+dy^2+dz^2$. 
Here, the coordinates ($N$,$\vec r$)
are dimensionless, the fifth coordinate
$\psi $ has spatial unities and $\epsilon$ is a dimensionless parameter
that can take the values $\epsilon = 1,-1$. 
The metric (\ref{6}) describes a
flat 5D manifold in apparent vacuum ($G_{AB}=0$).
We consider a diagonal metric because we are dealing only with
gravitational effects, which are the important ones in the global evolution
for the universe.
To describe neutral matter in a 5D geometrical vacuum
(\ref{6}) we can consider the Lagrangian
\begin{equation}\label{1}
^{(5)}{\rm L}(\varphi,\varphi_{,A}) =
-\sqrt{\left|\frac{^{(5)}
g}{^{(5)}g_0}\right|} \  ^{(5)}{\cal L}(\varphi,\varphi_{,A}),
\end{equation}
where $|^{(5)}g|=\psi^8 e^{6N}$
is the absolute value of the determinant for the 5D metric tensor with
components $g_{AB}$ ($A,B$ take the values $0,1,2,3,4$) and
$|^{(5)}g_0|=\psi^8_0 e^{6N_0}$
is a constant of dimensionalization determined
by $|^{(5)}g|$ evaluated at $\psi=\psi_0$ and $N=N_0$.
In this work we shall consider $N_0=0$, so that
$^{(5)}g_0=\psi^8_0$.
Here, the index ``$0$'' denotes the values at the end of inflation.
Furthermore, we shall consider an action
\begin{displaymath}
I = - {\Large\int} d^4x d\psi \sqrt{\left|\frac{^{(5)}g}{^{(5)}g_0}\right|}
\left[\frac{^{(5)} R}{16\pi G} + {\cal L}(\varphi,\varphi_{,A})\right],
\end{displaymath}
for a scalar field $\varphi$, which is minimally coupled to gravity.
Here, $^{(5)} R$ is the 5D Ricci scalar, which of course, is cero for
the 5D flat metric (\ref{6}) and $G$ is the gravitational constant.

Since the 5D metric (\ref{6}) describes a manifold in apparent
vacuum, the density Lagrangian
${\cal L}$ in (\ref{1}) must to be
\begin{equation}\label{1'}
^{(5)}{\cal L}(\varphi,\varphi_{,A}) = 
\frac{1}{2} g^{AB} \varphi_{,A} \varphi_{,B},
\end{equation}
which represents a free scalar field. In other words, we define the vacuum
as a purely kinetic 5D-lagrangian on a globally 5D-flat metric [in our
case, the metric (\ref{6})].
In the 3D comoving frame $U^r=0$,
the geodesic dynamics ${dU^C \over dS}=-\Gamma^C_{AB} U^A U^B$
with $g_{AB} U^A U^B=1$, give us the velocities $U^A$:
$U^{\psi} = - {1 \over \sqrt{u^2(N)-1}}$, $U^{r}=0$,
$U^N={u(N) \over \psi\sqrt{u^2(N)-1}}$,
which are satisfied for $S(N)=\pm |N|$.
In this work we shall consider the case $S(N) = |N|$.
In this representation ${d\psi \over dN}=\psi/u(N)$, where
$u(N)$ is an arbitrary function.
Thus the fifth coordinate evolves as
\begin{equation}\label{psi}
\psi(N) = \psi_0 e^{\int dN/u(N)}.
\end{equation}
Here, $\psi_0$ is a constant of integration that has spatial unities.
From the mathematical point of view, we are taking a foliation
of the 5D metric (\ref{6}) with $r$ constant.
Hence, to describe
the metric in physical coordinates we must to make the
following transformations:
$t = \int \psi(N) dN$, $R=r\psi$, $ L= \psi(N) \  e^{-\int dN/u(N)}$,
such that for $\psi(t)=1/h(t)$,
we obtain the 5D metric
\begin{equation}\label{m1}
dS^2 = \epsilon\left(dt^2 - e^{2\int h(t) dt} dR^2 - dL^2\right),
\end{equation}
where $L=\psi_0$ is a constant and $h(t)=\dot b/b$ is the effective
Hubble parameter defined from the effective scale factor of the
universe $b$.
The metric (\ref{m1}) describes a 5D generalized FRW
metric, which is 3D spatially flat [i.e., it is flat in terms of
$\vec R = (X, Y, Z)$], isotropic and homogeneous.
In the representation $(\vec R,t,L)$, the 
velocities $ \hat U^A ={\partial \hat x^A \over \partial x^B} U^B$,
are
\begin{equation} \label{10}
U^t=\frac{2u(t)}{\sqrt{u^2(t)-1}}, \qquad
U^R=-\frac{2r}{\sqrt{u^2(t)-1}}, \qquad U^L=0,
\end{equation}
where
the old velocities $U^B$ are $U^N$, $U^r=0$ and $U^{\psi}$
and the velocities $\hat U^B$ are constrained by the condition
\begin{equation}\label{con}
\hat g_{AB} \hat U^A \hat U^B =1.
\end{equation}
Furthermore, the function $u$ can be written as a function of time 
$u(t) = -{h^2 \over \dot h}$,
where the overdot represents the derivative with respect to the time.
The solution $N={\rm arctanh}[1/u(t)]$ corresponds to a
time dependent power-law expanding universe
$h(t)=p_1(t) t^{-1}$, such that the effective scale factor go as
$b \sim e^{\int p_1(t)/t dt}$.
When $u^2(t) >1$, the velocities $U^t$ and $U^R$ are real, so that
the condition (\ref{con}) implies that $\epsilon =1$. [Note that the
function $u(t)$ can be related to the deceleration parameter
$q(t) = -\ddot b b/\dot b^2$: $u(t) = 1/[1+q(t)]$.]
In such that case the expansion of the universe is accelerated ($\ddot b >0$).
However, when $u^2 <1$ the velocities $U^t$ and $U^R$ are imaginary
and the condition (\ref{con}) holds for $\epsilon = -1$.
In this case the expansion of the universe is decelerated because
$\ddot b <0$. So, the parameter $\epsilon$ is introduced in the metric
(\ref{m1}) to preserve the hyperbolic condition (\ref{con}).
Moreover, the coordinates $(\vec R,t,L)$ has physical meaning, because
$t$ is the cosmic time and $(\vec R,L)$ are spatial coordinates.
Since the line element is a function of time $t$
(i.e., $S\equiv S(t)$), the new coordinate $R$ give us
the physical distance between galaxies separated
by cosmological distances: $R(t)=r(t)/h(t)$.
Note that for $r >1$ ($r <1$), the 3D spatial distance $R(t)$ is defined
on super (sub) Hubble scales.
Furthermore $b(t)$ is the effective scale factor of the
universe and describes its effective 3D euclidean (spatial) volume (see
below).
Hence, the effective 4D metric
is a spatially (3D) flat FRW one
\begin{equation}\label{frw}
dS^2 \rightarrow ds^2 = \epsilon
\left(dt^2 - e^{2\int h(t) dt} dR^2\right),
\end{equation}
and has a effective
4D scalar curvature $^{(4)}{\cal R} = 6(\dot h + 2 h^2)$. The
metric (\ref{frw}) has a metric tensor with components $g_{\mu\nu}$
($\mu,\nu$ take the values $0,1,2,3$).
The absolute value of the determinant for this tensor is $|^{(4)}g|
=(b/b_0)^6$.
The density Lagrangian in this new frame was obtained in a previous
work\cite{MB}
\begin{equation}\label{aa}
^{(4)} {\cal L}\left[\varphi(\vec{R},t), \varphi_{,\mu}(\vec{R},t)\right]
= \frac{1}{2} g^{\mu\nu} \varphi_{,\mu} \varphi_{,\nu}\nonumber \\
 - \frac{1}{2} \left[(R h)^2  -
\frac{b^2_0}{b^2} \right] \  \left(\nabla_R \varphi\right)^2, 
\end{equation}
and the equation of motion for $\varphi$ yields
\begin{equation}\label{bb}
\ddot\varphi + 3 h\dot\varphi -\frac{b^2_0}{b^2} \nabla^2_R \varphi
+ \left[\left(4\frac{h^3}{\dot h} - 3\frac{\dot h}{h}
 - 
3\frac{h^5}{\dot h^2}\right) \dot\varphi +
\left( \frac{b^2_0}{b^2} - h^2 R^2\right)\nabla^2_R\varphi\right]=0.
\end{equation}
From eqs. (\ref{aa}) and (\ref{bb}), we obtain respectively
the effective scalar 4D potential
$V(\varphi)$ and its derivative with respect
to $\varphi(\vec{R},t)$ are
\begin{eqnarray}
V(\varphi) & \equiv & \frac{1}{2}\left[ (R h)^2
- \left(\frac{b_0}{b}\right)^2 \right] \left(\nabla_R\varphi\right)^2,
\label{au} \\
V'(\varphi)  \equiv  
 \left(4\frac{h^3}{\dot h} - 3\frac{\dot h}{h} -
3\frac{h^5}{\dot h^2}\right) \dot\varphi 
 + \left(\frac{b^2_0}{b^2} - h^2 R^2\right)\nabla^2_R\varphi,\label{a1}
\end{eqnarray}
where the prime denotes the derivative with respecto to $\varphi $.
The equations (\ref{aa}) and (\ref{bb}) describe the dynamics of the inflaton field
$\varphi(\vec{R},t)$ in a metric (\ref{frw}) with a Lagrangian
\begin{equation}\label{l4}
^{(4)}{\cal L}[\varphi(\vec{R},t),\varphi_{,\mu}(\vec{R},t)] =
-\sqrt{\left|\frac{^{(4)}g}{^{(4)}g_0}\right|}
\left[\frac{1}{2} g^{\mu\nu} \varphi_{,\mu}\varphi_{,\nu}
+V(\varphi)\right],
\end{equation}
where $\left|^{(4)}g_0\right|=1$.

Furthermore, the 4D energy 
density $\rho$ and the pressure ${\rm p}$ are\cite{PLB}
\begin{eqnarray}
&& 8 \pi G \rho = 3 h^2,\\
&& 8\pi G {\rm p} = -(3h^2 + 2 \dot h).
\end{eqnarray}
Note that
the function $u(t)$ can be related
to the deceleration parameter $q(t)=-\ddot b b/\dot b^2$: $u(t)=1/[1+q(t)]$.
From the condition (\ref{con}) we can differentiate
some different stages of the universe.
If
$u^2(t)={4 r^2 (b/b_0)^2 -1 \over 3} >1$, we obtain
that $r$ can take the values
$r > 1$ ($r < 1$) for
$b/b_0 < 1$ ($b/b_0 > 1$), respectively.
In this case $q < 0$, so that the expansion is accelerated.
On the other hand if
$u^2(t)={4 r^2 (b/b_0)^2 -1 \over 3} <1$, $r$ can take
the values $r< 1$ ($r > 1$) for
$b/b_0 > 1$ ($b/b_0 < 1$), respectively. In this
stage $q >0$ and the expansion of the universe is decelerated,
so that the function $u(t)$ take the values $0 < u(t) <1$
and the velocities (\ref{10}) become imaginary. Thus,
the metric (\ref{frw}) shifts its signature from $(+,-,-,-)$
to $(-,+,+,+)$.
When $u(t) =1$ the deceleration parameter
becomes zero because $\ddot b =0$.
At this moment the velocities (\ref{10})
rotates syncronically in the complex plane and
$r$ take the values $r=1$ or $r<1$, for $b/b_0=1$ or $b/b_0 >1$, respectively.

On the other hand, the effective 4D
energy density operator $\rho$ is
\begin{equation}
\rho = \frac{1}{2} 
\left[ \dot\varphi^2 + \frac{b^2_0}{b^2} \left(\nabla\varphi\right)^2
+2 V(\varphi)\right].
\end{equation}
Hence, the 4D expectation value of the Einstein equation
$\left<H^2\right> = {8\pi G \over 3} \left<\rho\right>$
on the 4D FRW metric (\ref{frw}), will be
\begin{equation}
\left<H^2\right> = \frac{4\pi G}{3} \left< \dot\varphi^2 +
\frac{b^2_0}{b^2} \left(\nabla\varphi\right)^2 + 2 V(\varphi)\right>,
\end{equation}
where $G$ is the gravitational constant and $\left<H^2\right>
\equiv h^2=\dot b^2/b^2$.
Now we can make a semiclassical treatment\cite{NPB}
for the effective 4D quantum field
$\varphi(\vec R,t)$, such that $<\varphi> = \phi_c(t)$:
\begin{equation}\label{semi}
\varphi(\vec{R},t) = \phi_c(t) + \phi(\vec{R},t).
\end{equation}
For consistence we take $<\phi>=0$ and $<\dot\phi>=0$.
With this approach the classical dynamics on the background 4D FRW
metric (\ref{frw}) is well described by the equations
\begin{eqnarray}
&& \ddot\phi_c + 3 \frac{\dot b}{b} \dot\phi_c + V'(\phi_c)=0, \label{ua} \\
&& H^2_c = \frac{8\pi G}{3} \left(\frac{\dot\phi^2_c}{2} + V(\phi_c)
\right),\label{hc}
\end{eqnarray}
where $H^2_c = \dot a^2/a^2$ and the prime denotes de derivative with
respect to the field.
In other words the scale factor $a$ only takes into account
the expansion due to the classical Hubble parameter, but
the effective scale factor $b$ takes into account both, clasical
and quantum contributions in the energy density: ${\dot b^2 \over b^2} =
{8\pi G \over 3} \left<\rho\right>$.
Since $\dot\phi_c=-{H'_c \over 4\pi G}$, from eq. (\ref{hc}) we obtain
the classical scalar potential $V(\phi_c)$ as a function of the
classical Hubble parameter $H_c$
\begin{displaymath}
V(\phi_c) = \frac{3 M^2_p }{8\pi} \left[ H^2_c - \frac{M^2_p}{12\pi}
\left(H'_c\right)^2 \right],
\end{displaymath}
where $M_p=G^{-1/2}$ is the Planckian mass.
The quantum dynamics is described by
\begin{eqnarray}
\left< H^2 \right> & = &  H^2_c + \frac{8\pi G}{3} 
\left< \frac{\dot\phi^2}{2} + \frac{b^2_0}{2 b^2} (\nabla\phi)^2 +
\sum_{n=1} \frac{1}{n!} V^{(n)}(\phi_c) \phi^n \right>.\label{fried} \\
\ddot\phi & + &  3 \frac{\dot b}{b}\phi - \frac{b^2_0}{b^2} \nabla^2\phi +
\sum_{n=1} \frac{1}{n!} V^{(n+1)}(\phi_c) \phi^n =0, \label{apr} 
\end{eqnarray}
In what follows we shall make the following identification:
\begin{equation}
\Lambda(t) = 8\pi G 
\left< \frac{\dot\phi^2}{2} + \frac{b^2_0}{2 b^2} (\nabla\phi)^2 +
\sum_{n=1} \frac{1}{n!} V^{(n)}(\phi_c) \phi^n \right>,
\end{equation}
such that
\begin{equation} \label{lam}
\frac{\dot b^2}{b^2} = \frac{\dot a^2}{a^2} + \frac{\Lambda}{3}.
\end{equation}
On cosmological scales, the fluctuations $\phi$ are small, so that
it is sufficient to make a linear approximation ($n=1$) for
the fluctuations. Thus, the second term in (\ref{lam})
is negligible on such that scales. However, the second
term in (\ref{lam}) could be important in the ultraviolet
spectrum and more exactly at Planckian scales. At these
scales the modes for $\phi$ should be coherent and the matter
inside these regions can be considered as dark.
Hence, the significative contribution for the
function $\Lambda(t)$ is given by
\begin{equation}\label{lam1}
\Lambda(t) \simeq 8\pi G 
\left.\left< \frac{\dot\phi^2}{2} + \frac{b^2_0}{2 b^2} (\nabla\phi)^2 +
\sum_{n=1} \frac{1}{n!} V^{(n)}(\phi_c) \phi^n \right>\right|_{Planck}.
\end{equation}
In this sense, we could make the identification for $\Lambda $ as
a cosmological parameter which only takes into account
the ``coherent quantum modes'' (or dark matter) contribution for the 
expectation value of energy density: $\left<\rho_{\Lambda}\right> =
\Lambda /(8\pi G)$. For simplicity, in the
following we shall consider $\Lambda$ as a constant.

Once done the linear approximation ($n=1$) for the semiclassical treatment
(\ref{semi}), we can make the identification of the squared mass for the
inflaton field $m^2 = V''(\phi_c)$\cite{PRD}. Hence, after make a linear expansion
for $V'(\varphi)$ in eq. (\ref{a1}), we obtain
\begin{eqnarray}
V'(\phi_c) & \equiv &
\left(4\frac{h^3}{\dot h} -
3\frac{\dot h}{h} - 3 \frac{h^5}{\dot h^2}\right)
\dot\phi_c,\label{ua1} \\
m^2 \phi & \equiv  &
\left(4 \frac{h^3}{\dot h} - 3 \frac{\dot h}{h} - 
3\frac{h^5}{\dot h^2} \right) \frac{\partial \phi}{\partial t} 
+ \left(\frac{b^2_0}{b^2} - h^2 R^2 \right) \nabla^2_R\phi.
\label{apr1}
\end{eqnarray}
Taking into account the expressions (\ref{ua}) with (\ref{ua1}) and
(\ref{apr}) with (\ref{apr1}), we obtain the dynamics for $\phi_c$ and $\phi$.
Hence, the equations $\ddot\phi_c+ 3h \dot\phi + V'(\phi_c)=0$
and $\ddot\phi +   3h\dot\phi_c - (b/b_0)^2\nabla^2_R \phi
+ V''(\phi_c)\phi=0$ now take the form\cite{MB}
\begin{eqnarray}
&& \ddot\phi_c + \left[3h+ f(t)\right]\dot\phi_c =0, \\
&&  \ddot\phi + \left[3h(t) + f(t)\right]\dot\phi -
h^2R^2\nabla^2_R\phi =0, \label{qf}
\end{eqnarray}
where
\begin{equation}\label{f}
f(t)= \left(4\frac{h^3}{\dot h} -
3\frac{\dot h}{h} - 3 \frac{h^5}{\dot h^2}\right).
\end{equation}

\section{An Example}

To ilustrate the formalism we consider a time dependent
power expansion $p(t) = 2/3 + A t^{-2} - B t^{-1}$, such
that the classical Hubble parameter is given
by $H_c(t)=p(t)/t$ and ($A$,$B$) are
constants.
The effective power $p_1(t)$
for the effective Hubble parameter $h(t)$ will be
$p_1(t) = \sqrt{(2/3 + A t^{-2} - B t^{-1})^2 + \Lambda/3 t^2}$,
because $h^2 = H^2_c + \Lambda/3$. In what follows we shall
consider the universe as spatially flat.
This implies that the total density parameter will be
$\Omega_T =\Omega_r + \Omega_m + \Omega_{\Lambda}=1$,
for a critical energy density given
by $\rho_c = {3 \over 8\pi G} h^2$, such that
\begin{equation}
\Omega_r + \Omega_m = \frac{H^2_c}{h^2}, \qquad \Omega_{\Lambda} = \frac{
\Lambda}{3 h^2} .
\end{equation}
where $\Omega_r$, $\Omega_m$ and $\Omega_{\Lambda}$ are respectively
the contributions for radiation, matter and $\Lambda$.
In our case, because we consider $\Omega_T = 1$, this implies that
\begin{equation}
p^2_1(t) = p^2(t) + \frac{1}{3} \Lambda t^2,
\end{equation}
where $t>0$ is the cosmic time.
We define $b/b_0 = e^N$, such that $b_0\equiv b(t=t_0)$, where
shall consider $t_0$ as the time when inflation ends (i.e., the
time for which $\ddot b=0$). Thus $N$ will be grater than zero
only for times larger than $t_0$, but negative for
$t< t_0$ (i.e., during the previous inflationary phase).
This means that the parameter $N$ give us
the number of e-folds with respect to the scale factor at the
end of inflation: $b_0$.
Once defined the scale for $N$, we
can see the evolution for the function $u(t)$.
During inflation $\ddot b >0$, so that $u(t) >1$
and $\epsilon = 1$. In such that
epoch $q <0$ (i.e., the universe is accelerated) and
$b/b_0=e^N < 1$, because $N <0$. In such that phase the parameter
$r$: $r \gg 1$. This means that cosmological
scales include regions very much larger than the Hubble horizon [see
the metric (\ref{frw})].

At the end of inflation $u(t)$ take values close (but
larger) to the unity. At $t=t_0$ $\ddot b=q=0$,
the function $u(t_0)=1$, so that
the global hyperbolic geometry condition
$\hat g_{AB} \hat U^A \hat U^B=1$
it's not well defined [see eqs. (\ref{10})].
However, the line element (\ref{frw}) is well
defined. At this
moment the universe suffers a change of phase from a accelerated 
to a decelerated expansion and $r=1$, because $b(t=t_0)=b_0$.

During the second phase (i.e., decelerated expansion) the universe
is governed by radiation and later by matter. The function
$u^2(t)$ is smaller than the unity (but $u^2 >0$), so that
$r$ take values
${1\over 2} e^{-N} ={1\over 2} b_0/b < r < 1$, for $N >0$.
This means that, during this phase, the metric (\ref{frw}) describes
the universe on scales smaller than the Hubble radius: $r/h < 1/h$.
The interesting here is that the velocities (\ref{10}) becomes
purely imaginary and the signature of the 4-D effective
metric (\ref{frw}) changes synhcronically
(with respect to the signature during the
inflationary phase): $(+,-,-,-) \rightarrow
(-,+,+,+)$; that is, $\epsilon$ jumps from $1$ to $-1$ to preserve
the global geometry in (\ref{con}).
In this sense we can say that the 4-D effective
metric (\ref{frw}) is ``dynamical''.
Note that this possibility was first considered by Davidson
and Owen in \cite{new}.
The fig. (\ref{gi231}) shows the evolution of the powers $p_1[x(t)]$
(dashed line) and
$p[x(t)]$ (continuous line) as a function of $x(t) = {\rm log}_{10}(t)$ for
$A=1.5 \  10^{30} \  {\rm G}^{1}$ and $B=10^{15} \  {\rm G}^{1/2}$.
Numerical calculations give us the time for which $\ddot b =q=0$
at the end of inflation: $x(t_0)\simeq
14.778$. At this moment $N(t_0)=0$, but after it becomes positive.
Note that for $x(t) < 60.22$ both curves are very similar, but for
$x(t) > x(t_*)$ (with $x(t_*) \simeq 60.22$), $p_1$
increases very rapidly but not $p$, which remains
almost constant with a value close to $p \simeq 2/3$. The difference
between both curves is due to the presence of
a nonzero ``cosmological constant'' ($\Lambda$),
which was valued as: $\Lambda= 1.5 \  10^{-121} \  {\rm G}^{-1}$.
[At the moment the consensus has
emerged about the experimental value of the cosmological constant\cite{u,u1}.
It is on the order of magnitude of the matter energy density:
$\rho_{\Lambda} \sim (2-3)\rho_{m}$.
The Wilkinson Microwave Anisotropy Probe (WMAP) data
suggest that the universe is very nearly spatially flat,
with a density parameter $\Omega_T = 1.02 \pm 0.02$\cite{spergel}.]
In other words, at $t_* \simeq 1.66 \  10^{60} \  {\rm G}^{1/2}$
the deceleration parameter becomes zero and later negative.
At this moment, the universe changes from a 
decelerated to a accelerated phase and $\epsilon$ jumps from
$-1$ to $1$ because $u(t)$ evolves from $u(t< t_*) <1$ (decelerated
expansion) to
$u(t> t_*) > 1$ (accelerated expansion). It should be when the universe
was nearly $0.4 \  10^{10}$ years old.
The present day age of the universe was considered as $x(t)
= 60.653 \  {\rm G}^{1/2}$ (i.e., $1.5 \  10^{10}$ years old).
Note that $\Omega_r+ \Omega_m$ decreases for late times
[see figure (\ref{gi232})], so that
its present day value should be
$(\Omega_r+ \Omega_m)[x(t=60.653 \  G^{1/2})] \simeq 0.32$.
Thus, the present day value for the vacuum density parameter
$\Omega_{\Lambda}=1-(\Omega_r+ \Omega_m)$ should be
$\Omega_{\Lambda}[x(t=60.653 \  G^{1/2})] \simeq 0.68$.
With these parameter values
we obtain the present day deceleration parameter:
$q[x(t=60.653 \  G^{1/2})] \simeq -0.747$, so that the present day
cosmological parameter should be: $\omega[x(t=60.653 \  G^{1/2})] \simeq
-0.831$. Note that all these results are in very good agreement with
observation\cite{PDG,spergel}.\\

\section{Final Comments}

The possibility that our universe is embedded in a higher dimensional
space has generated a great deal of active interest. In Brane-World and
STM theories the usual constraint on Kaluza-Klein models, namely the cylinder
condition, is relaxed so the extra dimensions are not restricted to be
compact. Although these theories have different physical motivations
for the introduction of a large dimension, they share the same
working scenario, lead to the same dynamics in 4D\cite{pdl}.
In this work we have studied a model for the evolution of the universe
which is globally described by a single scalar field from a 5D apparent
vacuum. Such that vacuum is described by the diagonal
metric (\ref{6}) and a
purely kinetic Lagrangian.
The 5D formalism here developed
could be extended to other particular frames or quantum fields.
Moreover, the
evolution of the universe could be examined taking into account
also electromagnetism by introducing off-diagonal terms in the
metric\cite{pdl1}, which should be relevant to study 3D spatial
anisotropies in the
universe on astrophysical scales.
However, all
these issues go beyond the scope of this work.

\begin{center}
\begin{figure}
\includegraphics[width=5cm, height=5cm]{c:/tex/gi231.bmp}
\vspace{-0cm}
\noindent
\caption{\label{gi231} Evolution of $p_1[x(t)]$ (dashed line) and $p[x(t)]$
(continuous line) as a function of $x(t) = {\rm log}_{10}(t)$, for
$A=1.5 \  10^{30} \  {\rm G}^{1}$, $B=10^{15} \  {\rm G}^{1/2}$.}
\end{figure}
\end{center}

\begin{center}
\begin{figure}
\includegraphics[width=5cm, height=5cm]{c:/tex/gi232.bmp}
\vspace{-0cm}
\noindent
\caption{\label{gi232} Evolution of $(\Omega_m + \Omega_r)[x(t)]$
as a function of $x(t) = {\rm log}_{10}(t)$, for
$A=1.5 \  10^{30} \  {\rm G}^{1}$, $B=10^{15} \  {\rm G}^{1/2}$.}
\end{figure}
\end{center}


\begin{thebibliography}{99}
\bibitem{jd} J. Dreitlein, Phys. Rev. Lett. {\bf 33}, 1243 (1974).
\bibitem{Guth} A. Guth, Phys. Rev. {\bf D23}, 347 (1981).
\bibitem{we} C. Wetterich, Nucl. Phys. {\bf B302}, 668 (1988);
P. J. E. Peebles and B. Ratra, Ap. J. Lett. {\bf 325}, L17 (1988);
R. R. Caldwell, R. Dave and P. J. Steinhardt, Phys. Rev. Lett. {\bf 80},
1582 (1998);
C. Armendariz-Picon, V. Mukhanov and P. J. Steinhardt, Phys. Rev. Lett.
{\bf 85}, 4438 (2000).
\bibitem{sta} T. D. Saini, S. Raychaudhurg, V. Shani and A. A. Starobinsky,
Phys. Rev. Lett. {\bf 85}, 1162 (2000).
\bibitem{wesson} J. M. Overduin, P. S. Wesson, Phys. Rep. {\bf 283},
303 (1997).
\bibitem{e1} P. S. Wesson and J. Ponce de Leon, J. Math. Phys. {\bf 33},
3883 (1992); P. S. Wesson, Mod. Phys. Lett. {\bf A7}, 921 (1992).
\bibitem{e2} P. S. Wesson, Astrophys. J. {\bf 394}, 19 (1992).
\bibitem{e3} A. P. Billyard and P. S. Wesson, Gen. Rel. Grav. {\bf 28},
129 (1996).
\bibitem{e31} J. E. Lidsey, C. Romero, R. Tavakol and S. Rippl,
Class. Quant. Grav. {\bf 14}, 865 (1997).
\bibitem{e4} B. Mashhoon, H. Liu and P. S. Wesson, Phys. Lett. {\bf B331},
305 (1994).
\bibitem{NPB} M. Bellini, Nucl. Phys. {\bf B604}, 441 (2001).
\bibitem{ps} I. Zlatev, Li-Min Wang and P. J. Steinhardt,
Phys. Rev. Lett. {\bf 82}, 896 (1999).
\bibitem{PLB} D. S. Ledesma and M. Bellini, Phys. Lett. {\bf B581}, 1 (2004).
\bibitem{MB}
J. E. Madriz Aguilar and M. Bellini,
Phys. Lett. {\bf B596}, 116 (2004).
\bibitem{PDG} Review of Particle Physics, Phys. Rev. {\bf D66}, 010001-170
(2002).
\bibitem{PRD} M. Bellini, H. Casini, R. Montemayor and P. Sisterna,
Phys. Rev. {\bf D54}, 7172 (1996).
\bibitem{new} A. Davidson and D. Owen, Phys. Lett. {\bf B177}, 77 (1986).
\bibitem{u} T. Padmanabhan, Phys. Rep. {\bf 380}, 235 (2003).
\bibitem{u1} D. N. Spegel, L. Verde, H. V. Peiris, {\em et al}.,
Ap. J. Suppl. {\bf 148}, 175 (2003).
\bibitem{spergel} D. N. Spergel {\em et al.,} Astrophys. J. Suppl. {\bf 148},
175 (2003).
\bibitem{pdl} J. Ponce de Leon, Mod. Phys. Lett. {\bf A16}, 2291 (2001).
\bibitem{pdl1} J. Ponce de Leon, Int. J. Mod. Phys. {\bf D11}, 1355 (2002).
\end{thebibliography}
\end{document}